\documentclass[showkeys,showpacs]{revtex4} 
\usepackage{amsmath} 
\usepackage{amssymb} 
\usepackage{graphicx} 
\usepackage{color} 

\begin{document} 

\title{Nonperturbative quantum corrections} 
\author{Vladimir Dzhunushaliev} 
\email{vdzhunus@krsu.edu.kg} 
\affiliation{Dept. Phys. and Microel. Engineer., Kyrgyz-Russian Slavic University, Bishkek, Kievskaya Str. 44, 720021, Kyrgyz Republic} 

\date{\today}

\begin{abstract}
A nonperturbative quantization procedure based on a nonassociative decomposition of quantum field operators on nonassociative constituents is considered. It is shown that such approach gives rise to quantum corrections by calculations of expectation values of nonlinear functions of field operators. The corrections can in principle be measured as a radius of a force, characteristic length of nonlocal objects, the failure of connection compatibility with metric, and so on. The system of gravity interacting with Maxwell electromagnetism is considered. It is shown that quantum corrections from gravitoelectric coupling of a certain form leads to vanishing singularities of a point charge, including infinite self-energy. 
\end{abstract}


\maketitle

\section{Introduction} 

One of the most serious problems in modern physics is the quantization of strongly interacting quantum fields. This includes the confinement problem in quantum chromodynamics, quantization of gravitation, and probably also high temperature superconductivity with its strong interaction between Cooper electrons. The problem is that the algebra of quantum operators describing strongly interacting fields is unknown. Known commutation relationships of type
\begin{equation}
\label{1-10}
  \left[ \hat \phi(x), \hat \phi(y) \right] = i \delta(x-y)
\end{equation}
describe \textcolor{blue}{\emph{free, noninteracting fields}} (here $ \hat \phi (x) $ is the operator of a free field $ \phi (x) $). 

The need for nonperturbative techniques in strongly interacting, nonlinear quantum field theories is an old problem that has been around since the beginning of the study of quantum fields. Much effort has gone into trying to resolve  this puzzle. The different approaches that have been tried include: (i) lattice QCD  \cite{Teper:1998kw}, (ii) the dual Meissner effect in the QCD-vacuum \cite{Pandey}-\cite{Agasian}, (iii) instantons \cite{Suganuma} \cite{Shifman},  (iv) path integration \cite{Kondo5}, (v) analytic calculations \cite{Simonov}, (vi) Dyson-Schwinger equations \cite{Gentles}. Despite this, the problem is not yet fully resolved. All these approaches are approximations ones. 

In the 1950's, Heisenberg \cite{Heisenberg1} \cite{Heisenberg2} studied a nonlinear spinor field and developed nonperturbative techniques for quantizing the nonlinear spinor field. His method assumes an infinite system of equations which relate all the n-point Green's functions of the theory (this can be compared to the infinite number of Feynman diagrams which must, in principle, be calculated for a given process in pertubative quantum field theory). In order to solve this system of equations, one must find some physically reasonable cut-off approximations, so that one reduces the infinite system of equations into a finite system. Nevertheless, the algebra of field operators in Heisenberg's approach remains unknown. 

In Ref. \cite{coleman2} it was shown that radiative corrections could introduce a symmetry breaking ({\it i.e.} negative) mass term into a scalar Lagrangian. This effect is called dimensional transmutation. Similar reasonings in Ref. \cite{Zloshchastiev:2009aw} give rise to that the logarithmic nonlinearity in the quantum wave equation can cause the spontaneous symmetry breaking 
and mass generation phenomena. One can presuppose that a \textcolor{blue}{\emph{nonperturbative quantization}} of any strongly interacting fields would yield similar terms. In Ref. \cite{Dzhunushaliev:2009na-dec} it is offered to quantize strongly interacting fields using nonassociative (n/a) decomposition of the field into products of n/a factors. In such approach the rearrangement of brackets gives rise to additional terms in the same way as the permutation of field operators (in standard quantum field theory) gives rise to the Planck constant. 

\section{Nonassociative decomposition of quantum field operators}

In this section we follow to Ref. \cite{Dzhunushaliev:2009na-dec}. We assume that operators of strongly interacting fields $\Phi_m \left (x^\mu \right)$ can be decomposed into n/a constituents $f^i_\alpha$ and $b_{i \beta}$:
\begin{equation}
\label{2-10}
  \Phi_m \left( x^\mu \right) = f^i_\alpha \left( x^\mu \right)
  b_{i \beta} \left( x^\mu \right)
\end{equation}
here $m$ is an index where internal and Lorentzian indices are collected, $i$ is the summation index, and the $\alpha$, $\beta$ are contained in $m$ as: $m = \left \{\alpha, \beta \right \}$. Although the constituent operators $f^i_\alpha, b _ {i \beta}$ are not associative, the basic idea requires their product to model an associative operator. In more mathematical terms, the operators $f^i_\alpha$ and $b _ {i \beta}$ are elements in a n/a algebra 
$\mathbb A $, i.e., $f^i_\alpha , b _ {i \beta} \in \mathbb A \setminus \mathbb G$, which contains an associative subalgebra $\mathbb G \subset \mathbb A $, such that $\Phi_m  = f^i_\alpha b_{i \beta} \in \mathbb G$. It is necessary to note that the operator $\Phi_m$ models observable quantities, whereas the n/a $f^i_\alpha, b _ {i \beta}$ are unobservable. 

Let us note that the number of n/a constituents does not need to be two, but may be more. For example Eq. \eqref{2-10} can be rewritten in the form 
\begin{equation}
\label{2-20}
  \Phi_m \left( x^\mu \right) = \left( 
  	q^{\phantom{1}}_{1\alpha} \left( x^\mu \right)
  \right)^i_j
  \left( 
  	q^{\phantom{2}}_{2\alpha} \left( x^\mu \right)
  \right)^j_k
  \left( 
  	q^{\phantom{3}}_{3\gamma} \left( x^\mu \right)
  \right)^k_i
\end{equation}
here $i,j,k$ are the summation indices, and the $\alpha$, $\beta$ and $\gamma$ are contained in $m$ as: $m = \left \{\alpha, \beta, \gamma \right \}$. One can say that the decompositions \eqref{2-10} and \eqref{2-20} somewhat correspond  to slave-boson \cite{slaveboson} and spin-charge (or quark-like) \cite{spincharge} decompositions. 

\section{Applications}

In this section we would like to consider a few examples: scalar field theory with polynomial $\phi^4$, pure gravity, and gravity interacting with electromagnetic field. 

\subsection{Scalar field theory with strong self-interaction}
\label{scalar}

Let us consider scalar field theory with Lagrangian 
\begin{equation}
\label{3-20}
  \mathcal L = \frac{1}{2} \nabla^\mu \phi \nabla_\mu \phi - V(\phi)
\end{equation}
where the nonlinear potential term is 
\begin{equation}
\label{3-25}
  V(\phi) = \frac{\lambda}{4} \phi^4 \left( x^\mu \right) .
\end{equation}
We assume a n/a decomposition 
$\phi \left( x^\mu \right) = f^{i} \left( x^\mu \right) b_{i} \left( x^\mu \right)$. Here, the operator $\phi$ is an observable associative quantity, but $f^{i}, b_{i}$ are unobservable n/a quantities. Using n/a factors one can rewrite the potential term from Lagrangian \eqref{3-20} as follows 
\begin{equation}
\label{3-30}
  \biggl(f^{i_1} \left( x^\mu \right) b_{i_1} \left( x^\mu \right) \biggr)
  \biggl(f^{i_2} \left( x^\mu \right) b_{i_2} \left( x^\mu \right) \biggr)
  \biggl(f^{i_3} \left( x^\mu \right) b_{i_3} \left( x^\mu \right) \biggr)
  \biggl(f^{i_4} \left( x^\mu \right) b_{i_4} \left( x^\mu \right) \biggr).
\end{equation}
In order to calculate an expectation value we have to define the action of operators on a quantum state. We will require: 
\begin{equation}
\label{3-40}
  \phi \left| \psi \right\rangle = \left( f^{i} b_{i} \right) \left| \psi \right\rangle 
  \stackrel{def}{=} 
  f^{i} \left( b_{i} \left| \psi \right\rangle \right) .
\end{equation}
This is the same rule as for the associative case. But when having two 
or more nonassociative operators, there is: 
\begin{equation}
\begin{split}
\label{3-50}
  \phi^2 \left| \psi \right\rangle = & \Bigl( 
  	\left( f^{i} b_{i} \right) \left( f^{i} b_{i} \right) 
  \Bigr) \left| \psi \right\rangle = \biggl( \Bigl( 
	  \left( f^{i} b_{i} \right)  f^{i}
	\Bigr)  b_{i} \biggr) \left| \psi \right\rangle + \mathrm{assoc} \left| \psi \right\rangle = 
	\Bigl( 
	  \left( f^{i} b_{i} \right)  f^{i}
	\Bigr)  \left( b_{i} \left| \psi \right\rangle \right) 
	 + \mathrm{assoc} \left| \psi \right\rangle = 
\\
	& 
	\left(f^{i} b_{i} \right) \Bigl( \bigl( f^{i}
	\left( b_{i} \left| \psi \right\rangle \right) \bigr) \Bigr)
	 + \mathrm{assoc} \left| \psi \right\rangle = 	
	f^{i} \Bigl( b_{i} \bigl( f^{i}
	\left( b_{i} \left| \psi \right\rangle \right) \bigr) \Bigr)
	 + \mathrm{assoc} \left| \psi \right\rangle 
\end{split}
\end{equation}
where the associator Ass is defined as follows 
\begin{equation}
\label{3-60}
 	\left( f^{i} b_{i} \right) \left( f^{i} b_{i} \right) = \biggl( \Bigl( 
	  \left( f^{i} b_{i} \right) f^{i}
	\Bigr)  b_{i} \biggr) + \mathrm{assoc} .
\end{equation}
In order to define the associator, $assoc$, we recall that in the standard commutations relationship \eqref{1-10} on the LHS we have two operators (observables) and on the RHS we have $2-2=0$ operators. In a similar way, we assume that on the RHS of \eqref{3-60} should be $2-2=0$ \textcolor{blue}{\emph{associative}} operators. It means that $\mathrm{assoc} = m^2$ where $m$ can be a complex number. One can say that it is the second Planck constant but with different dimension. 

The same can be done for $\phi^3 \left| \psi \right\rangle$
\begin{equation}
\label{3-70}
  \phi^3 \left| \psi \right\rangle = 
	f^{i} \Biggl( b_{i} \biggl( f^{i} \Bigl( b_{i} \bigl( f^{i}
	\left( b_{i} \left| \psi \right\rangle \right) \bigr) \Bigr) \biggr) \Biggr)
	 + m^2 \phi \left| \psi \right\rangle 
\end{equation}
and for $\phi^4 \left| \psi \right\rangle$
\begin{equation}
\label{3-80}
  \phi^4 \left| \psi \right\rangle = 
	f^{i} \left( b_{i} \left(
		f^{i} \Biggl( b_{i} \biggl( f^{i} \Bigl( b_{i} \bigl( f^{i}
		\left( b_{i} \left| \psi \right\rangle \right) \bigr) \Bigr) \biggr) \Biggr)
	\right) \right) 
	 + m^2 \phi^2 \left| \psi \right\rangle .
\end{equation}
Then the expectation value will be 
\begin{eqnarray}
\label{3-90}
  \left\langle \psi \left| \phi^3 \right| \psi \right\rangle &=& \left\langle \psi \left| 
		f^{i} \left( b_{i} \left( f^{i} \left( b_{i} \left( f^{i}
		\left( b_{i} \right| \psi \right\rangle \right) \right) \right) \right) \right)
	 + m^2 \left\langle \psi \left| \phi \right| \psi \right\rangle ,
\\
\label{3-100}
  \left\langle \psi \left| \phi^4 \right| \psi \right\rangle &=& \left\langle \psi \left| 
	f^{i} \left( b_{i} \left(
		f^{i} \left( b_{i} \left( f^{i} \left( b_{i} \left( f^{i}
		\left( b_{i} \right| \psi \right\rangle \right) \right) \right) \right) \right)
	\right) \right) 
	 + m^2 \left\langle \psi \left| \phi^2 \right| \psi \right\rangle .
\end{eqnarray}
Ordinarily, the vacuum expectation value 
$\left\langle \psi \left| \phi \right| \psi \right\rangle = \left\langle \phi \right\rangle$ of a field is zero. Consequently, in the case where $\left| \psi \right\rangle$ is a vacuum state, there is 
\begin{equation}
\label{3-110}
  \left\langle \psi \left| \phi \right| \psi \right\rangle = 0.
\end{equation}
It means that the n/a terms appear with the potential $V(\phi) \propto \phi^n, n\geq 4$ only. 

Thus the conclusion for this section is that \textcolor{blue}{\emph{the nonassociative corrections appear by calculation of an expectation value in nonperturbative quantum field theory for strongly enough nonlinear terms in the Lagrangian.}}

\subsection{Gravity}
\label{gravity}

General relativity is the most nonlinear theory we know. Consequently nonperturbative effects must be very important in quantum gravity. In general relativity the first nonlinearity is connected with a connection $\left\{^{\alpha}_{\beta \gamma}\right\}$ compatible with a metric 
\begin{equation}
\label{4-10}
	\left\{^{\alpha}_{\beta \gamma}\right\} = \frac{1}{2} g^{\alpha \delta} \left(
		g_{\beta \delta ,\gamma} + g_{\gamma \delta ,\beta} - 
		g_{\beta \gamma ,\delta}
	\right)
\end{equation}
which is called as Christoffel symbols. The nonlinearity is created by the quantity $g^{\alpha \delta}$ which is the matrix reciprocal to the matrix $g_{\alpha \delta}$. Such nonlinearity is much stronger than the potential term \eqref{3-20} in usual quantum field theory. How $\hat{g}^{\alpha \delta}$ can be described as an operator in nonperturbative quantization is unclear. In this section we denote operators as $\widehat{\ldots}$. Nevertheless, we continue to carry out the idea about appearance of additional terms from calculation of an expectational value. For the Christoffel symbols in this case we will have 
\begin{equation}
\label{4-20}
	\left\langle \psi \left| 
		\widehat{\left\{^{\alpha}_{\beta \gamma}\right\}} 
	\right| \psi \right\rangle = \left\langle 
		\left\{^{\alpha}_{\beta \gamma}\right\}
	\right\rangle = \widetilde{
	\left\{^{\alpha}_{\beta \gamma}\right\}} + 
	K_{\beta \gamma}^{\phantom{\beta \gamma} \alpha} = 
	\Gamma_{\beta \gamma}^{\phantom{\beta \gamma} \alpha}
\end{equation}
here $\left\langle \ldots \right\rangle$ is quantum averaging; 
$\widetilde{ \left\{^{\alpha}_{\beta \gamma}\right\}} = 
\Gamma_{(\beta \gamma)}^{\phantom{\beta \gamma} \alpha}$ is the symmetric part of the affine connection 
$\Gamma_{\beta \gamma}^{\phantom{\beta \gamma} \alpha}$; 
$K_{\beta \gamma}^{\phantom{\beta \gamma} \alpha} = 
\Gamma_{[\beta \gamma]}^{\phantom{\beta \gamma} \alpha}$ is the antisymmetric part of 
$\Gamma_{\beta \gamma}^{\phantom{\beta \gamma} \alpha}$ and describes the deviation quantum Christoffel symbols 
$\left\langle \left\{^{\alpha}_{\beta \gamma}\right\}	\right\rangle$ from the classical 
$\widetilde{ \left\{^{\alpha}_{\beta \gamma}\right\}}$; 
$\Gamma_{\beta \gamma}^{\phantom{\beta \gamma} \alpha}$ is the designation for 
$\left\langle \left\{^{\alpha}_{\beta \gamma}\right\}	\right\rangle$. The interesting question here is: why quantum averaging of the symmetric affine connection $\widetilde{ \left\{^{\alpha}_{\beta \gamma}\right\}}$ leads to a non-symmetric affine connection 
$\Gamma_{\beta \gamma}^{\phantom{\beta \gamma} \alpha}$ ? In our opinion it happens because the wave functional $| \psi \rangle$ is non-symmetric one. 

For the first approximation we assume that 
\begin{equation}
\label{4-30}
	\left\langle 
		\left\{^{\alpha}_{\beta \gamma}\right\} \left\{^{\mu}_{\nu \rho}\right\} 
	\right\rangle 
	\approx  
	\left\langle 
		\left\{^{\alpha}_{\beta \gamma}\right\} 
	\right\rangle 
	\left\langle 
		\left\{^{\mu}_{\nu \rho}\right\} 
	\right\rangle .
\end{equation}
In this case the quantum averaged Ricci tensor is 
\begin{equation}
\label{4-40}
	\left\langle R_{\mu \nu} \right\rangle \approx 
	\frac{\partial \Gamma_{\mu \nu}^{\phantom{\mu \nu} \rho}}
	{x^{\rho}} - 
	\frac{\partial \Gamma_{\mu \rho}^{\phantom{\mu \rho} \rho}}
	{x^{\nu}} + 
	\Gamma_{\mu \nu}^{\phantom{\mu \nu} \rho} \Gamma_{\rho \tau}^{\phantom{\mu \tau} \tau} - 
	\Gamma_{\mu \rho}^{\phantom{\mu \rho} \tau} \Gamma_{\nu \tau}^{\phantom{\mu \tau} \rho}.
\end{equation}
It means that we can think about the quantum averaged Ricci tensor 
$\left\langle R_{\mu \nu}(\left\{ \right\}) \right\rangle$ as about the Ricci tensor $\mathcal R_{\mu \nu}(\Gamma)$ with a connection $\Gamma_{\beta \gamma}^{\phantom{\beta \gamma} \alpha}$ which is not-compatible with the metric $g_{\mu \nu}$. 

Let us remember some notions from the differential geometry. In general the affine connection $\Gamma_{\beta \gamma}^{\alpha}$ can be written as 
\begin{equation}
\label{4-50}
	\Gamma_{\mu \nu}^{\phantom{\mu \nu}\rho} = 
	\left\{^{\rho}_{\mu \nu}\right\} + 
	K_{\mu \nu}^{\phantom{\mu \nu}\rho} 
\end{equation}
where $\left\{^{\alpha}_{\beta \gamma}\right\}$ are the usual Christoffel symbols of the symmetric connection and the contortion tensor $K$ is called the contorsion tensor and is given in terms of the torsion tensor by 
\begin{eqnarray}
\label{4-60}
	K_{\mu \nu}^{\phantom{\mu \nu}\rho} &=& \frac{1}{2} g^{\rho \sigma}
	\left(
		T_{\mu \sigma \nu} + T_{\nu \sigma \mu} - T_{\mu \nu \sigma}
	\right) ,
\\
\label{4-70}
	K_{[\mu \nu ]}^{\phantom{[ \mu \nu ]}\rho} &=& - \frac{1}{2} 
	T_{\mu \nu}^{\phantom{\mu \nu}\rho},
\\
\label{4-80}
	K_{\mu \nu \rho} &=& - K_{\mu \rho \nu}
\end{eqnarray}
here the torsion tensor $T_{\mu \nu}^{\phantom{\mu \nu}\rho}$ is the antisymmetric part of the affine connection coefficients $\Gamma_{\mu \nu}^{\phantom{\mu \nu}\rho}$. 
\begin{equation}
\label{4-90}
	T_{\mu \nu}^{\phantom{\mu \nu}\rho} = -2 
	\Gamma_{[\mu \nu ]}^{\phantom{[ \mu \nu ]}\rho}
\end{equation}
According to \eqref{4-40} and definitions \eqref{4-60} -- \eqref{4-80} we can say that the connection in classical general relativity (Christoffel symbols) after nonperturbative quantization acquires an additional contribution in the form of torsion. As well one can say that the torsion appears in quantum gravity as the result of nonperturbative quantization. 

\subsection{Gravity coupled with electrodynamics}

Above we have shown that by nonperturbative quantization of gravity the torsion appears as a quantum correction to Christoffel symbols. It means that the Einstein -- Cartan gravity is the first approximation for quantum gravity. In classical and perturbative quantum electrodynamics there is a big problem with an infinite energy of static electric field created by a point charge. One hope is that quantum gravity resolves this problem.

\subsubsection{Quantum corrections from gravitational nonlinearities}
\label{ed_gravity_1}

In this subsection we would like to show that above mentioned quantum corrections smooth this problem. These corrections should be taking into account on a small distance only. For simplicity, we consider a metric with torsion only (but otherwise flat). We can then consider Maxwell electrodynamics in Minkowski spacetime (following to Ref. \cite{Ponomarev:1978zy} we use $(-,+,+,+)$ signature here and in the next subsection). The Lagrangian in this case is 
\begin{equation}
\label{4-1-10}
	\mathcal L = \frac{1}{16 \pi c} F_{\mu \nu}\left( \Gamma \right) 
	F^{\mu \nu}\left( \Gamma \right) 
\end{equation}
where 
\begin{equation}
\label{4-1-20}
	F_{\mu \nu}\left( \Gamma \right) = \nabla_\mu A_\nu - \nabla_\nu A_\mu = 
	\partial_\mu A_\nu - \partial_\nu A_\mu - T_{\mu \nu}^{\phantom{\mu \nu} \rho} A_\rho 
\end{equation}
here $A_\rho$ is 4-potential of the electromagnetic field and $F_{\mu \nu}\left( \Gamma \right)$ is corresponding tensor of electromagnetic field for the affine connection $\Gamma$. Let us note that the electromagnetic tensor \eqref{4-1-20} is not gauge invariant in the consequence of the torsion term $T_{\mu \nu}^{\phantom{\mu \nu} \rho}$. Maxwell equations can be written in the form \cite{Ponomarev:1978zy} 
\begin{eqnarray}
\label{4-1-30}
	\frac{1}{\sqrt{-g}} \frac{\partial }{\partial x^\mu} 
	\left( 
		\sqrt{-g} F^{\mu \nu}\left( \Gamma \right)
	\right)	&=& \frac{4 \pi}{c} J^\nu , 
\\	
\label{4-1-35}
	F_{\mu \nu}\left( \Gamma \right) &=& \mathcal F_{\mu \nu} + 
	\frac{2 G}{c^4} \frac{A_{[ \mu} F_{\nu ] \rho} A^\rho}
	{1 + \frac{G}{c^4} A_\mu A^\mu}
\\	
\label{4-1-40}
	J^\nu &=& - \frac{G}{4 \pi c^3} F^{\mu \nu}\left( \Gamma \right) 
	F_{\mu \rho}\left( \Gamma \right) A^\rho 
\end{eqnarray}
here $\mathcal F_{\mu \nu} = \partial_\mu A_\nu - \partial_\nu A_\mu$ and $G$ is Newton constant. 
Allowing quantum corrections from torsion, we now consider an electrostatic spherically symmetric 
solution with 4-potential 
\begin{equation}
\label{4-1-50}
	A_\mu = \left(
		\phi (r), 0,0,0
	\right).
\end{equation}
Then Eq. \eqref{4-1-30} has the form 
\begin{eqnarray}
\label{4-1-60}
	\mathrm{div} \vec E &=& 4 \pi \rho, 
\\	
\label{4-1-70}
	\rho &=& \frac{G}{4 \pi c^4} \vec E^2 \phi , 
\\	
\label{4-1-80}
	E_r &=& F_{0r}\left( \Gamma \right) = 
	\frac{\mathcal F_{0r}}{1 + \frac{G}{c^4} \phi^2}.
\end{eqnarray}
The spherical solution is 
\begin{eqnarray}
\label{4-1-90}
	\phi &=& \frac{c^2}{\sqrt{G}} \sinh \left(
		\frac{q \sqrt{G}}{c^2} \frac{1}{r}  
	\right),
\\	
\label{4-1-100}
	E_r &=& \frac{q}{\cosh \left( 
		\frac{q \sqrt{G}}{c^2} \frac{1}{r}
	\right)} \frac{1}{r^2} , 
\\	
\label{4-1-110}
	\rho &=& \frac{\sqrt{G}}{4 \pi c^2} 
	\frac{\tanh \left( \frac{q \sqrt{G}}{c^2} \frac{1}{r} \right)}
	{\cosh\left( \frac{q \sqrt{G}}{c^2} \frac{1}{r} \right)} 
	\frac{q^2}{r^4} 
\end{eqnarray}
where $q = \int \limits_V \rho d^3 x$ is the electric charge. Interestringly, the electric field $E_r$ and charge density $\rho$ are nonsingular at the origin 
\begin{equation}
\label{4-1-120}
	E_r(0) = \rho(0) = 0. 
\end{equation}
However, the total energy of the field becomes infinite:
\begin{equation}
\label{4-1-130}
	\int \limits_V \left(
		\frac{1}{8 \pi} \vec E^2 - \frac{1}{2} \rho \phi
	\right) d^3 x = \infty .
\end{equation}
While the integral $\int \limits_V \vec E^2 d^3 x < \infty$ is finite, the total energy becomes 
infinite due to the self-interaction term $\rho \phi$. Evidently, the coupling between electrodynamics and gravity is nonlinear. Consequently, one can hope that including nonperturbative corrections for the electromagnetic field connected with gravity will give rise to a finite energy of the electric field of a point charge.

\subsubsection{Quantum corrections from nonlinear gravitoelectric coupling}
\label{ed_gravity_2}

Quantum corrections coming from the interaction between electromagnetic and gravitational fields are unknown in the case of very strong nonlinear gravitoelectric coupling. We assume that the corrections can be written as $- \frac{d V(A_\mu)}{d A_\nu}$ in following form in Maxwell equations 
\begin{equation}
\label{4-2-10}
	\frac{1}{\sqrt{-g}} \frac{\partial }{\partial x^\mu} 
	\left( 
		\sqrt{-g} F^{\mu \nu}\left( \Gamma \right)
	\right)	= \frac{4 \pi}{c} J^\nu - \frac{d V(A_\mu)}{d A_\nu}
\end{equation}
where $V(A_\mu)$ are quantum nonperturbative corrections for the electromagnetic potential in the consequence of nonlinear system: gravity + electromagnetism. These corrections cannot be calculated in general, but for some $V(A_\mu)$ there exist \textcolor{blue}{\emph{regular}} solutions. 

\begin{figure}[h]
\begin{minipage}[t]{.45\linewidth}
  \begin{center}
  \fbox{
  \includegraphics[width=.8\linewidth]{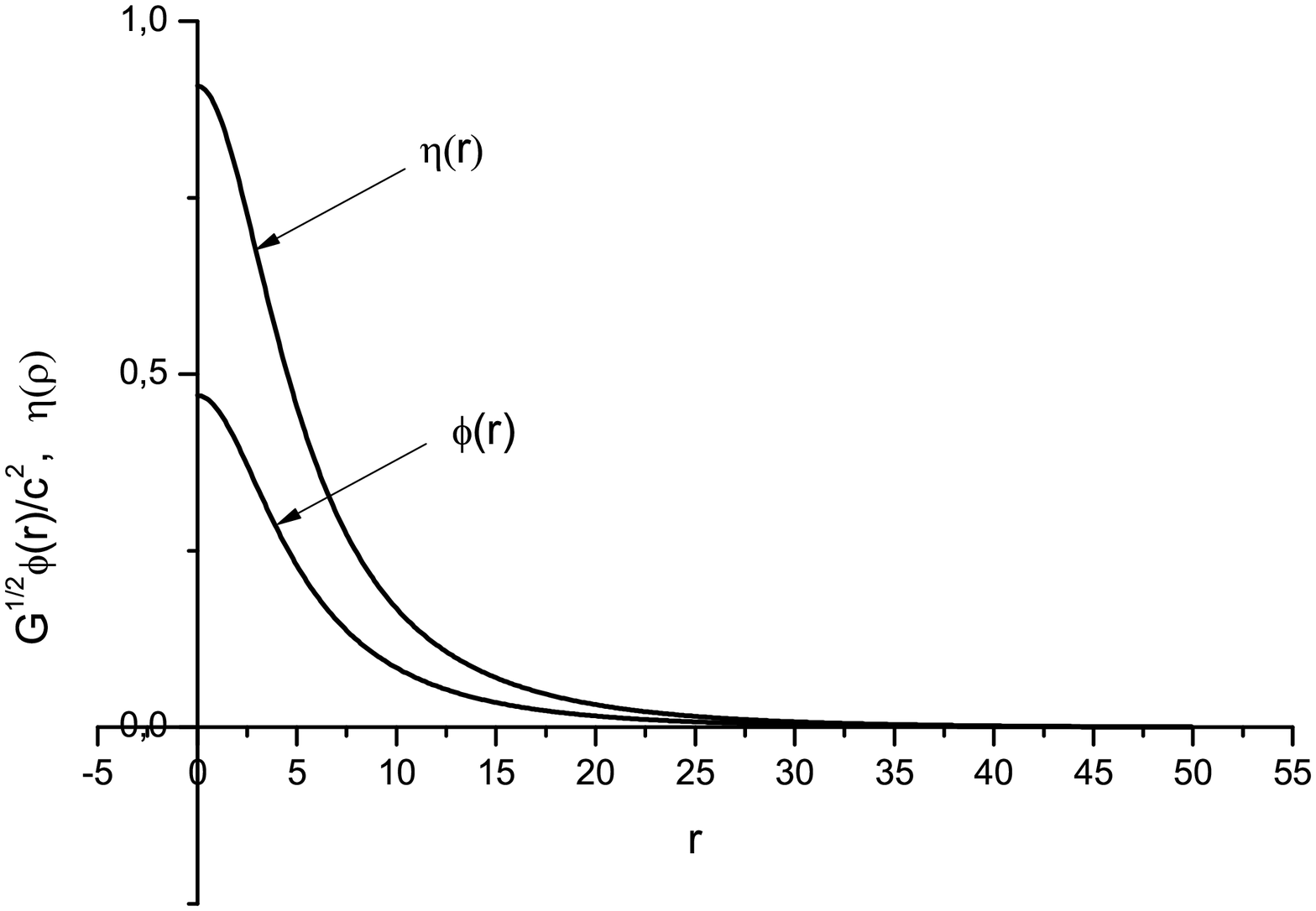}}
  \caption{The profiles of $\eta(r)$ and $\frac{\sqrt{G}}{c^2} \phi(r)$. 
  $\tilde \lambda = 1, m = 0.1, \eta^*_0 = .9083$.}
  \label{eta}
  \end{center}
\end{minipage}\hfill
\begin{minipage}[t]{.45\linewidth}
  \begin{center}
  \fbox{
  \includegraphics[width=.8\linewidth]{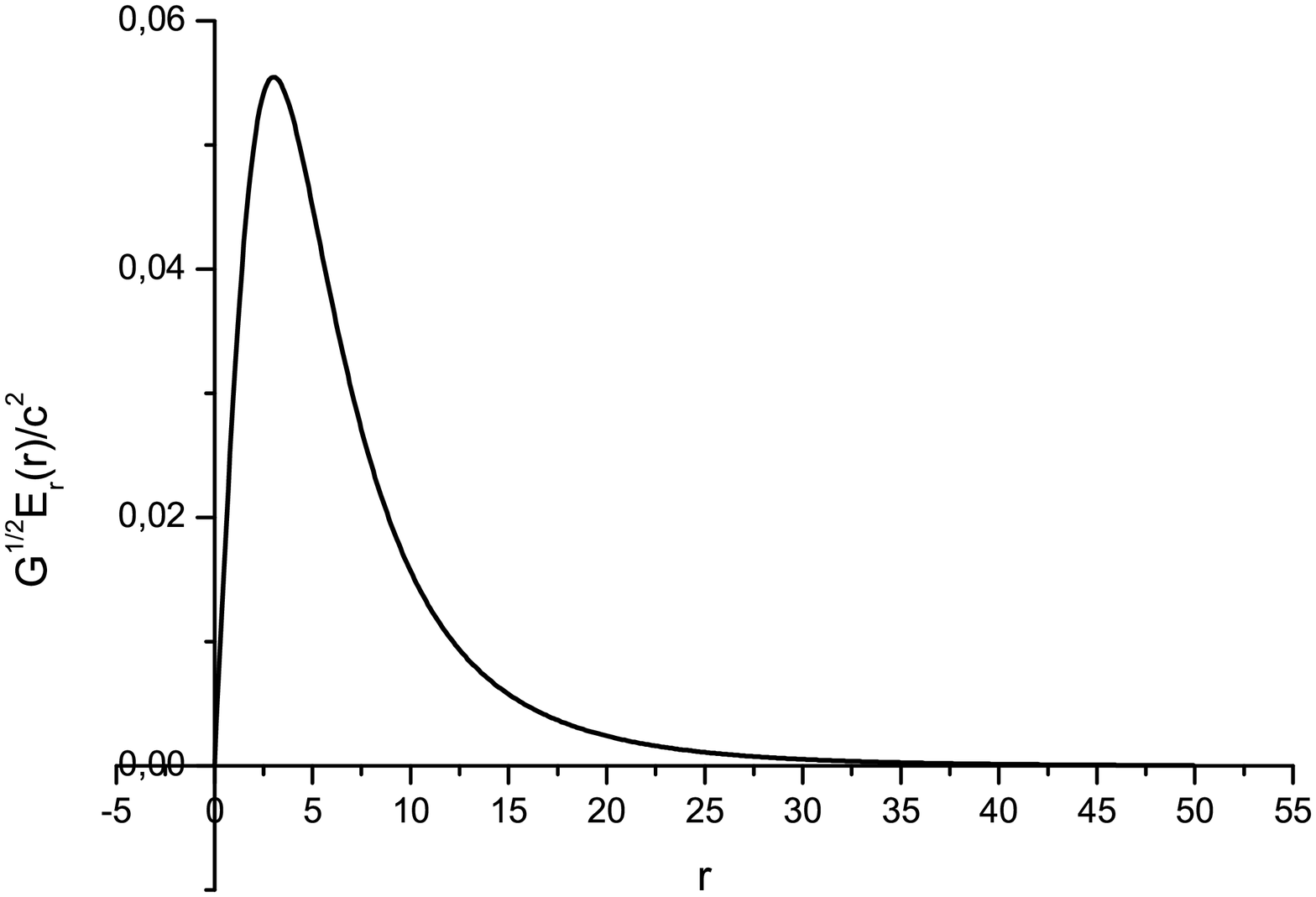}}
  \caption{The profile of the electric field $\frac{\sqrt{G}}{c^2} E_r(r)$. 
  $\tilde \lambda = 1, m = 0.1, \eta^*_0 = .9083$.}
  \label{electric}
  \end{center}
\end{minipage}\hfill
\end{figure}

\begin{figure}[ht]
\begin{minipage}[t]{.45\linewidth}
  \begin{center}
  \fbox{
  \includegraphics[width=.8\linewidth]{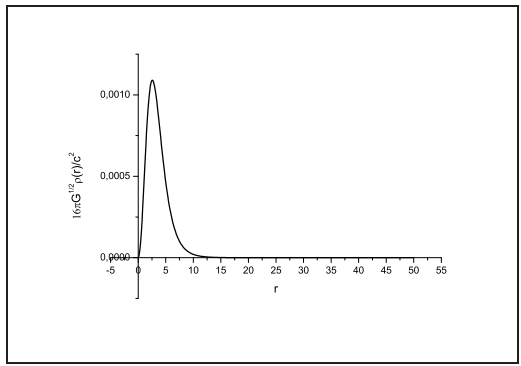}}
  \caption{The profile of the charge density $16 \pi \frac{\sqrt{G}}{c^2} \rho(r)$. 
  $\tilde \lambda = 1, m = 0.1, \eta^*_0 = .9083$.}
  \label{rho}
  \end{center}
\end{minipage}\hfill
\end{figure}

We now assume quantum corrections of the form 
$V(A_\mu) = \frac{\lambda}{4} \left( A_\mu A^\mu + A_0^2 \right)^2$. For the spherically symmetric 4-potential \eqref{4-1-50} Maxwell equation is 
\begin{equation}
\label{4-2-20}
	\frac{1}{r^2} \left( r^2 \eta^\prime \right)^\prime	= 
	-\tilde \lambda \sinh \eta \left[
		\sinh^2 \left( \frac{\eta}{2} \right) - m^2
	\right] 
\end{equation}
where $\phi(r) = \frac{c^2}{\sqrt{G}} \sinh\left[ \frac{\eta(r)}{2} \right]$; 
$\tilde \lambda = \frac{c^4}{G} \lambda$ and $m^2 = \frac{G}{c^4} A_0^2$. We are searching for the regular solution at the origin. Consequently the boundary conditions are
\begin{eqnarray}
\label{4-2-25}
	\eta(0) &=& \eta_0, \\
\label{4-2-27}
	\eta^\prime(0) &=& 0. 
\end{eqnarray}
We doubt that there exists an analytical solution of Eq.\eqref{4-2-20}. Therefore we are searching for a numerical solution. The numerical investigation shows that a special regular solution $\eta^*(r)$ does exist for some special choice of $\eta(0) = \eta^*_0$ only. The results of numerical solution of Eq.\eqref{4-2-20} in Fig's \ref{eta} - \ref{rho} are presented. Since the solution is special, this makes mass, charge, and other physical properties of such an electric charge unique.

\section{How one can measure nonassociativity in physics}

The question stated in the title of this section is analogous to standard quantum theory: How can one measure \textcolor{blue}{\emph{the noncommutativity}} of operators? For the standard quantum theory, the answer is: The Planck constant measures the noncommutativity of conjugated operators.

In this paper, we proposed an approach to the nonperturbative quantization through nonassociative decomposition of quantum field operators. During a lunch with Geoffrey Dixon, Tevian Dray, John Huerta, Jens K\"oplinger and Shahn Majid (on 2$^{nd}$ Mile High Conference on Nonassociative Mathematics, Denver, Colorado, USA) the question came up, on how one could measure nonassociativity in physics ? 

We have considered two examples: (1) polynomial potential in Section \ref{scalar}; (2) gravity in \ref{gravity}. The calculations presented in \ref{scalar} for the scalar fields can be extended to any field theory having a polynomial potential term (for example, for a gauge theory). In gauge theory the Lagrangian is 
\begin{equation}
\label{5-10}
	\mathcal L = \frac{1}{4} F^a_{\mu \nu} F^{a \mu \nu}.
\end{equation}
where $F^a_{\mu \nu} = \partial_\mu A^a_\nu - \partial_\nu A^a_\mu + g f^{abc} A^b_\mu A^c_\nu$ is the field strength; $A^a_\mu$ is the gauge potential; $a,b,c = 1, \ldots ,N$ are the SU(N) color indices; $g$ is the coupling constant; $f^{abc}$ are the structure constants for the SU(N) gauge group. 

The Lagrangian \eqref{5-10} has the term $f^{abc} f^{ade} A^b_\mu A^c_\nu A^{d \mu} A^{e \nu}$ which is similar to the scalar potential $\phi^4$. Reasoning similar to section \ref{scalar} leads to appearance of an additional mass term $m^2 A^a_\mu A^{a \mu}$. Such mass term controls the radius of the interaction $A_\mu (r) \sim e^{- m r}/r$. It means that the nonassociativity manifests itself as the radius of the interaction $r_{int} \sim 1/m$. The appearance of the mass term $m^2 A^a_\mu A^{a \mu}$ signifies breaking of gauge invariance. Quantum chromodynamics is a 
field theory with strong interaction, for which above arguments are applicable. The radius $r_{int} \sim 1/m$ can be considered as a radius of a flux tube filled with a chromoelectric field and stretched between quark - antiquark. Thus n/a parameter $m^{-1}$ can be measured in principle. 

In Section \ref{gravity} we have considered quantum corrections for gravity. Exact calculations cannot be done in this case because we do not know the exact form of the nonperturbative operators $\hat g^{\mu \nu}$, $\widehat{\sqrt{-g}}$, and so on. We have proposed the torsion as a quantum correction from the Christoffel symbols. In subsection \ref{ed_gravity_1}1 we have shown that such corrections lead to smoothing of singularities, together with infinite total field energy of a point charge. In subsection \ref{ed_gravity_2} we have considered the nonlinear system of gravity + electrodynamics. In such system we also can not calculate quantum corrections. But we have shown that if the quantum correction has some definite form then all infinities in point charge disappear. 

In summarizing, nonassociative quantum corrections can be measured: 
\begin{itemize}
	\item In nonperturbative quantum field theory, nonperturbative (nonassociative) quantum corrections can be measured as a radius of corresponding forces.
	\item In gauge theories with big enough coupling constant, the nonassociativity gives rise to breaking of gauge invariance and formation of nonlocal objects (flux tubes) with characteristic length reciprocal to n/a parameter $m$.
	\item In pure gravity, n/a quantum corrections appear in affine connections as torsion. It can in principle be measured, but probably only on very small (Planck) distances.
	\item In gravity + electromagnetism, system n/a quantum corrections probably gives rise to smoothing of all singularities connected with a point charge. It leads to the possibility to modeling of a zero-spin charged particles. 
\end{itemize}

\section{Discussion and conclusions}

Here we have considered nonperturbative quantization procedures based on nonassociative decomposition of quantum field operators, into nonassociative constituents. We have seen that such approach gives rise to quantum corrections, by calculation of expectation values of field 
operators on nonlinear functions \footnote{potential term $V(\phi) = \phi^n, n \geq 4$ in a scalar field theory; $f^{abc} f^{ade} A^b_\mu A^c_\nu A^{d \mu} A^{e \nu}$ in a gauge theory, Christoffel symbols in gravity}. We have shown that these corrections can in principle be measured as a radius of a force, characteristic length of nonlocal objects, and the failure of connection compatibility with metric. Also in such way one can regularize singularities of a point-like charge. Let us note that in Ref. \cite{Zaslavskii:2010yi} similar idea was considered. It has been shown that in non-linear electrodynamics in the framework of general relativity there exist ``weakly singular'' configurations such that (i) the proper mass is finite in spite of divergences of the energy density, (ii) all field and energy distributions are concentrated in the core region. 

In Section \ref{ed_gravity_2} we have shown that quantum corrections having a false vacuum and two true vacuums gives rise to regular solution describing a regular charge distribution. The corrections considered allow us to assume that such peculiarity will be retained for nonperturbative quantum corrections similar to a Mexican hat potential.

According to \eqref{4-1-80} the torsion is controlled by the factor 
$\frac{G}{c^4} \phi^2 = \sinh(\eta/2)$. If $ \sinh(\eta/2) \ll 1$ then quantum corrections can be neglected. Otherwise, the corrections join the solutions for electric Coulomb fields. Such construction may model an isolated electric charge in Minkowski spacetime. 

Quantum mechanics manifests itself through appearance of the Planck constant $\hbar$. In this paper, we showed that nonassociativity may exhibit a constant in quantum field theory, $m^2$; 
or otherwise become evident as a geometric property in quantum gravity, torsion. Further investigation is be needed to clarify these relations. 

\section{Acknowledgments} 

I am grateful to the Research Group Linkage Program of the Alexander von Humboldt Foundation for financial support, and to Geoffrey Dixon, Tevian Dray, John Huerta, Jens K\"oplinger and Shahn Majid for fruitful discussions.  Special thanks to F.-W. Hehl for the fruitful discussion and criticism.

\end{document}